\documentclass[12pt]{article}

\begin{document}
\vskip 1cm

\centerline{ A Note on the Set of After-meaurement States }
\centerline { in Generalized Quantum Measurement}
\vskip 0.3cm
\centerline{ I.D. Ivanovic }
\centerline{ Physics Department,  Carleton University}
\centerline{Ottawa ON. Canada}
\centerline {igor@physics.carleton.ca }
\vskip 0.3cm
{\bf Abstract} The sets of after-measurement states for standard and generalized quantum measurements are compared.
It is shown that for a SIC-POVM generalized measurement, the ratio of the volume of the set of after-measurement states and the volume of the simplex generated by individual outcoms  quckly tends to zero with increase of the number of dimensions. The volumes used are based on the Hilber-Schmidt norm. Some consequences on actual realizations,  having  finite collections of systems are discussed.  

   .

\vskip 0.5cm
 {\bf States and Standard Measurements}. Few well known facts about standard measurements may be useful.  Let the system be described in a d-dimensional complex Hilbert space $H$. A pure state is a projector onto a normalized vector  $|\Psi\rangle , \langle \Psi|\Psi \rangle  =1 $, and its  state is a ray projector $P=|\Psi\rangle  \langle  \Psi|$. Generally, a state is an operator on $H $ satisfying  $W\geq0, trW=1 $. A state can be  expressed as a weighted sum of its eigen-projectors  $W=\sum _kr_kP_k$ where $\sum _kr_k=1$ , and  $P_kP_r=\delta _{kr}P_k$. The set of all states over $H$ will be denoted by $V_W=\{W|W\ge 0,tr(W)=1\}$.   
An orthogonal ray -resolution of the identity (ORRI) over  $H$ is a set of projectors $\{P_k\}$ satisfying 
$$ \sum _kP_k = I_H \quad \quad , P_kP_r =\delta _{kr} P_k \quad \quad  tr(P_k) =1 $$ 
The convex set of all convex combinations of an ORRI $\{P_k\}$,  $conv(\{P_k\})$  is a commutative simplex,  identical to the classical, discrete probability simplex, having pure states $P_k$ as extremal points and normalized identity  $W_o={1 \over d} I_H$ as its baricenter.  
The set  of all states, $V_W$,  can be obtained by applying all unitary transformations to  an initial, commutative,  simplex. The point  common to all simplices is $W_o$.\par
The most natural way to look at $V_W$ is as a convex set in the space of Hermitian operators over $ H $ using Hilbert-Schmidt distance .  A standard measurement , defined by   complete, nondegenerate observable $A $, having ORRI  $\{P_k\}$  is represented by a change of state  
$$ W_{am}=\sum_k P_kW_{pm}P_k=\sum_k tr(WP_k)P_k \eqno(1) $$
where  $p(a_k)=tr(W_{pm}P_k)$ and $\langle A\rangle =\sum _kp(a_k)a_k=tr(AW_{pm})=tr(AW_{am})$.
Here $W_{pm}$ is a pre-measurement state and $W_{am}$ is the after-measurement state of the system,or to be more precise of an infinite ensemble of systems. \par 
  Formally ,  $W_{am}$  is an orthogonal projection of $W_{pm}$ onto the simplex defined by 
the $\{P_k\}$. Again, the easiest way to visualize  this is to deduct the $W_o$ from all states, working in the hyperplane $tr(A)=1$,   
then the simplex of commuting states defined by  $\{P_k\}$ is in. e.g. two dimensions is a segment of length $\sqrt{2}$. The midpoint is $W_o$.   In three dimensions a commutative  simplex is equilateral triangle, edge  $\sqrt {2}$, the baricenter is, as always, $W_o$.\par

    Once an ORRI $\{P_k\}$ is given one may identify the three set of states:\par
i)  set of $tr=1$  linear combinations of $P_k$'s , $$V(\{P_k)\}) = \{W| W= \sum a_kP_k, \sum a_k =1,W\ge 0\},$$ 
ii) set of all convex combinations of $P_k$'s, 
$$ conv(\{P_k\})=\{W|W=\sum _ka_kP_k, \quad a_k\ge 0,\quad \sum _ka_k=1\}$$
iii)  the set  of all possible aftermeasurement states 
$$V_{am}(\{P_k\}) = \{W| W=\sum _kP_kW_{pm}P_k, W_{pm}\in V_W\}.$$ 
All   three sets are identical 
$$ V(\{P_k\})= conv(\{P_k\})= V_{am}(\{P_k\}).$$     
Furthermore, this type of measurement, corresponding to an ORRI , can be selective e.g. when the systems are 'tagged' and states with outcome $a_k$ are selected.  The other type is non-selective when $W_{am}$ is all we know about the state.  ORRI  measurements are  repeatable i.e. immediately after e.g. $ a_k$ is observed on a system, another measurement of observable A should give the same result and a consequence   it that it is also repeatable on the ensemble i.e. $\sum _kP_kW_{am}P_k=W_{am}$.
 Obviously, almost no measurements satisfy these conditions but as a paradigm it mirrors  our ideas of distinguishability  into orthogonality.\par
\vskip 1cm
{\bf Generalized Quantum Measurement}  In a generalized quantum measurement  (GM) the resolution of the identity is as a rule  a nonorthogonal one (NRI) and it is formally given by 
$$ W_{am}=\sum _kA_kW_{pm}A_k^{\dagger }$$
where $\sum _kA_k^{\dagger }A_k =I_d$ [1] but it is possible that $\sum _kA_kA_k^{\dagger}\neq I_d $. The possibility for a GM to displace $W_o$ indicates that there is a part of it which is a preparation, not  simply  a measurement . 
To make  things simple we will consider only non-orthogonal ray resolutions (NRRI), which  will be 'stripped '  of their  unitary part.  Namely,  using the polar decomposition  $A_k=U_kQ_k$   only ray-projector factor  $Q_k$ will be kept . The subset of GM we will consider is then   
$$  W_{am} = \sum  _k  c_k  Q_k  W_{pm} Q_k , \quad \sum _k  c_k  Q_k  = I, \quad tr(Q_k )=1 .$$
where $Q_k$s are a linearly independent set. In this way, a GM is always a contraction  having $W_o$ as 
one of the fixed points.\par

Realizations of GM , come as a rule,   come from a Naimark-like  constructions, either by expanding the space, making $H=H_S$ a subspace of a larger space, ${\cal H}=H_S\oplus H_A$, or by making H a factor space of ${\cal H} = H_S\otimes H_A$. \par
  In the first case[2],  the original space $H_S$  is  enlarged so that the NRRI  $\{Q_k\}$  is a projection of an ORRI from the enlarged space i.e. $c_kQ_k = P_SP_kP_S$ where $\{P_k\}$ is an ORRI from the enlarged space $\cal H$ and $P_S$ is the projector onto original space H. One must notice that the measurement should be made with $\{P_r\}$s on a state from $ V_W$ and then projected or rotated back into $ V_W$.\par

  A more frequent situation is when an ancila is attached to the system. In this case, an ORI performed on the ancila, after a unitary transformation is performed on $H_S\otimes H_A$,  results in an NRI measurement  on the system. The most straightforwrd construction is given in [3 ].\par
\hskip 1cm   
The system in state $W_{pm}$ is attached to the ancila in a specified state e.g. $P_o^A$. A unitary transformation is then applied to the $W_{pm}\otimes P_o^A$
 resulting in $U_{S\otimes A}W_{pm}\otimes P_o^AU_{S\otimes A}^{\dagger}$ such that 
$$ W_{am} =tr_A(\sum _k(I\otimes P_k^A) (U_{S\otimes A}W_{pm}\otimes P_o^AU_{S\otimes A}^{\dagger}) (I\otimes P_k^A)=\sum _kc_k tr(W_{pm}Q_k)Q_k\eqno (2)$$. 

 One should notice that the measurement of $I_S\otimes \{P_k^A\}$ on the ancila serves only to tag the systems in $H$ while the state of the system is already  the one given by eq.(2). So one needs a classical communication between the ancila and the system to identify individual systems and their states. Strictly speaking, no actual measurement is performed on the system,  what happened is an unitary transformation and a 'distant' selection'  [4].  The state of the system, after the unitary transformation is already
$$ tr_A(U_{S\otimes A}W_{pm}\otimes P_o^AU_{S\otimes A} ^{\dagger }) =tr_A(\sum _k(I_S\otimes P_k^A) (U_{S\otimes A}W_{pm}\otimes P_o^AU_{S\otimes A}^{\dagger}) (I_S\otimes P_k^A)$$

{\bf What is a measurement result in a GM ?}    In the case of a GM based on an NRRI   $ \{Q_k\})$,satisfying $\sum _k c_kQ_k=I, trQ_k=1$, NRRI  defines three sets of states:\par   
i) set of all $tr =1$ linear combinations of $\{Q_k\}$
$$V(\{Q_k\}) =\{ W | W=\sum a_kQ_k\ge 0, \quad \sum a_k=1, \quad a_k  -  real   \}$$
ii)  noncommutative simplex    $$ conv( \{Q_k\})=\{ W | W=\sum a_kQ_k\ge 0\quad , \quad a_k\ge 0, \quad  \sum a_k=1 \}$$  
and \par
iii)  the set of all possible after-measurement states  
$$V_{am}(\{Q_k\})= \{ W| W= \sum _kc_kQ_kW_{pm}Q_k, W_{pm}\in V_W \} $$ \par
 It is easy to see that 
$$  V(\{Q_k\})\supset   conv( \{Q_k\}) \supset  V_{am}(\{ Q_k\})$$      
  The first inclusion is obvious, the second follows if one  performs a measurement on  one of the extremal points from   $conv(\{Q_k\}) $, e.g. $W_{pm}=Q_{k_o}$. The state after the measurement is 
$$ W_{am} = \sum _k c_ktr(Q_kQ_{k_o})Q_k= c_{k_o}Q_{k_o} + \sum _{k\ne  k_o}c_ktr(Q_{k_o}Q_k)Q_k $$
In order for $Q_{k_o}$ to remain an extremal point of $conv(\{Q_k\}) $ , $c_{k_o}$ must be 1 and $tr(Q_kQ_{k_o}) =0$. \par
Therefore. due to nonorthogonality between  the ray-projectors from  $\{Q_k\}$, and in this case the lack of repeatability, the map of at least some  of the extremal points of $ conv(\{Q_k\})$ can not  remain  extremal points,  otherwise this NRRI would be an ORRI. \par  
 As commented  in [5], if the result of a measurement on certain number of identically prepared systems is still outside of   $V_{am}(\{Q_k\})$ , one should continue with measurements till the after -measurement state touches the boundary of  $V_{am}(\{Q_k\})$  
or goes into  $V_{am}(\{Q_k\})$. Should one continue with measrement or stop at the boundary ? This, of course, has no bearing on  an infinite ensemble, but it may affect any actual realization.\par 
 An interesting situation may occur in the following situation. Assume that we know nothing about $W_{pm}$, while the resulting  $W_{am}$,  after certain finite number of observations,  is still outside  of  $V_{am}(\{Q_k\})$: one may be forced to change the expected values of a subset of states for the second part of the ensemble, knowing that the final result should belong to  $V_{am}(\{Q_k\})$, or to be prepared to say that quantum mechanical description is incomplete.  Furthermore an observer on S  may communicate the results to ancila A , making the future results for an ORI on the ancila also more predictable.\par
Finally, what is actually measured?  In principle,  one can calculate  the expected values of all observables which are a linear combinations of $\{Q_k\}$; also, depending on the span of projectors, a position of a pre-measrement state is
reduced to a better defined subset of $V_W$. \par
\vskip 0.3cm
   {\bf SIC-POVM }   If an NRRI is symmetric-informationally complete SIC-POVM [6 ] i.e. if $d^2$ projectors $\{Q_k\}$ satisfy 
$${1 \over d} \sum _k Q_k =I_H \quad ; \quad  tr(Q_kQ_r)={{d\delta _{kr} +1}\over {(d+1)}}$$
one can make some more specific  conclusions.\par
 First, $\{Q_k\}$ spans the operator space and $V(\{Q_k\})\supset V_W$. 
This means that  any  pre-measurement state may be written  as $W_{pm}= \sum _k a_kQ_k $. The after-measurement state is then  
$$W{am}= \sum _{k,r}a_kc_r tr(Q_kQ_r)Q_r={1\over {(d+1)}} I + {1\over {(d+1)}}W_{pm}=$$

 $$  =  {d \over {(d+1)}}W_o+{1\over {(d+1)}}W_{pm}$$ 
 So,  in this measurement all states are contracted ( in the $ tr(A)=1$ hyperplane)
by a factor of  ${1\over {(d+1)}}$ (cf. [7]) . First thing that one may observe is that all after-measurement states must be  nonsingular. One can say that unless all events from $\{Q_k\}$ occur the state is definitely not allowed as a result.\par
   Furthermore,  the set of states "shrinks", but  the original shape of  $V_W^{(d)}$ is preserved .  A possible problem  is that we do not have a simple characterization or parameterization of the set of states, so even if an after-measurement state is inside the  sphere of radius ${1\over {(d+1)}}\sqrt {{(d-1)}\over d}$, it may not be an image of a state, rather, one would have to "stretch" the state to its original size to establish was  it actually a state or not.\par
Finally, the set of admissible after-measurement states  shrinks really quckly with incresed d. Due to the fact that all three sets 
$$ V_W  \supset conv(\{Q_k\}) \supset V_{am}(\{Q_k\}) $$  
have the same dimensions , one can compare their volumes.  \par 
The volume  
  of the $conv(\{Q_k\})$ in the hyperplane $tr(A)=1$, which is a $d^2-1$ dimensional simplex of edge $\displaystyle \Bigg ({{2d}\over {d+1}}\Bigg )^{1/2}$   is 
$$ {\cal V} (conv \{Q_k\}) = \Bigg ( {{2d}\over {(d+1)}}\Bigg )^{\displaystyle {{(d^2-1)}\over 2}}{d\over {\Bigg ( (d^2-1) !\quad 2^{\displaystyle {{(d^2-1)}\over 2}}\Bigg )}}.$$
The volume of states is , cf. [8],   
$${\cal V } (V_W) = \sqrt d   \Big (2\pi\Big )^{\displaystyle {{d(d-1)}\over 2}}  {{\Gamma (1)\dots \Gamma (d)}\over {\Gamma (d^2)}} $$
and the volume of the after-measurement states ( results ) for a SIC-POVM  $\{Q_k\}$ is then 
$$ {\cal V }(V_{am}(\{Q_k\}) = {{{\cal V}(V_W)} \over {(d+1)^{(d^2-1)}}}$$
 As a result, almost imediately, even for small d's    
$$ {{{\cal V}(conv(\{Q_k\}) )}\over {{\cal V} (V_W) }} \rightarrow   0 $$
but also 
$$ {{{\cal V}(V_{am} )}\over {{\cal V} (conv \{Q_k\}) }} \rightarrow   0 $$
Again, for infinite ensembles this is unimportant, but for any actual realization it probably is.\par 
  One should also notice that in a state reconstruction, tomography, or state determination, when it is made using ORIs a similar but not as drastic situation may occur. E.g.  first ORI measurement fixes a set of admissible pre-measurement states, the result of the following measurements must fit into it. The simplest  situation would be if the result of e.g. meassurement of spin 1/2 component  $S_z$ gives distribution $\{1-a,a\}$. If the next measuremesnt is e.g. of $S_x$ than as long as the result is outside   $\{1/2+b,1/2- b\}$  where  $ -\sqrt{a(1-a)} \leq b \leq\sqrt{a(1-a)} $ , the result of the state determination is  actually not a state.\par
    To conclude with, generalized measurements are indeed generalization of standard ORI measurements, but when they are not ORIs or combinations of ORIs  they are mostly either clever state determinations or distant state preparations. It is indeed very difficult to change a well established name, as generalized measurement is, but  more specifications may be  necessary. \par        
{\bf NB}  A part of this note was presented in poster session during  CAP Congress June 2010, Toronto,Canada. 
\vskip 1cm 
\leftline {\bf References:}
[1] K. Kraus.{\it States, Effects, and Operations} (Springer-Verlag, 1983 )\par 
[2] e.g. C.W. Helstrom:{\it Quantum Detection and Estimation Theory }  (Academic Press, 1976 )\par
[3] M.A. Nielsen and  I.L.Chuang, {\it QuantumComputation and Quantum Information}, Cambridge University Press (2000) \par
[4] F. Herbut and M. Vujicic, {\it Ann.Phys},{\bf 96},(1976)p382 \par
[5] A. Peres and   D.R. Terno; arXive:  quant-ph/9806024  \par
[6] e.g. A. Klappenecker {\it at all}:arXiv:quant-ph/0503239 \par 
[7] J. A. Rosado: arXive:1007.0715[quant-ph]\par
[8]  K. Zyczkowski and H.J. Sommers: arXive quant-ph/0302197\par

\end{document}